# Solvent Exfoliation of Electronic-Grade, Two-Dimensional Black Phosphorus

*Joohoon Kang[1], Joshua D. Wood[1], Spencer A. Wells[1], Jae-Hyeok Lee[1], Xiaolong Liu[2], Kan-Sheng Chen[1], and Mark C. Hersam[1,2,3,4]\**

[1]Department of Materials Science and Engineering, Northwestern University, Evanston, IL 60208
[2]Graduate Program in Applied Physics, Northwestern University, Evanston, IL 60208
[3]Department of Chemistry, Northwestern University, Evanston, IL 60208
[4]Department of Medicine, Northwestern University, Evanston, IL 60208
\* Correspondence should be addressed to m-hersam@northwestern.edu

**ABSTRACT**

Solution dispersions of two-dimensional (2D) black phosphorus (BP) – often referred to as phosphorene – are achieved by solvent exfoliation. These pristine, electronic-grade BP dispersions are produced with anhydrous, organic solvents in a sealed tip ultrasonication system, which circumvents BP degradation that would otherwise occur *via* solvated $O_2$ or $H_2O$. Among conventional solvents, n-methyl-pyrrolidone (NMP) is found to provide stable, highly concentrated (~0.4 mg/mL) BP dispersions. Atomic force microscopy, scanning electron microscopy, transmission electron microscopy, Raman spectroscopy, and X-ray photoelectron spectroscopy show that the structure and chemistry of solvent-exfoliated BP nanosheets are comparable to mechanically exfoliated BP flakes. Additionally, residual NMP from the liquid-phase processing suppresses the rate of BP oxidation in ambient conditions. Solvent-exfoliated BP nanosheet field-effect transistors (FETs) exhibit ambipolar behavior with current on/off ratios and mobilities up to ~$10^4$ and ~50 $cm^2V^{-1}s^{-1}$, respectively.  Overall, this study shows that stable, highly concentrated, electronic-grade 2D BP dispersions can be realized by scalable solvent exfoliation, thereby presenting opportunities for large-area, high-performance BP device applications.

Black phosphorus (BP),[1,2] a layered, anisotropic[3,4] allotrope of phosphorus, is emerging as a successor to other two-dimensional (2D) nanomaterials such as graphene[5,6] and transition metal dichalcogenides (TMDs)[7-9] due to its exceptional electronic properties. Unlike semi-metallic graphene, BP is a semiconductor with a thickness-dependent, direct band gap ranging from ~0.3 eV in the bulk to ~1.5 eV in the monolayer (*i.e.*, phosphorene) limit.[10-14] Mechanically exfoliated 2D BP possesses current on/off ratios[12, 15] of ~$10^4$-$10^5$ and room temperature mobilities up to ~200-1000 $cm^2V^{-1}s^{-1}$.[4, 12, 15-17] These desirable electronic properties make 2D BP a promising candidate for high-performance electronic and optoelectronic device applications.

Many production methods for 2D nanomaterials have been developed including micromechanical exfoliation,[5, 12, 18-24] chemical vapor deposition,[25-28] chemical exfoliation,[29,30] and liquid-phase exfoliation.[6, 8, 31-35] While micromechanical exfoliation generally produces 2D nanomaterials with the highest crystal quality, this method has limited scalability. Conversely, chemical vapor deposition is an effective method for thin-film applications, assuming that suitable precursors can be identified. Large quantities of 2D nanomaterials can also be produced by chemical exfoliation (*e.g.*, lithium intercalation[29] and oxidation[36,37]), but this approach typically introduces defects and/or leads to phase transformations that compromise electronic properties. Alternatively, liquid-phase exfoliation *via* ultrasonication[8] or shear mixing[33] are viable options to prepare 2D nanomaterials without intermediate chemical reactions.

Recently, it was reported that mechanically exfoliated 2D BP flakes irreversibly degrade to oxidized phosphorus compounds following ambient exposure.[18, 23, 38] This chemical instability presents challenges for BP liquid-phase exfoliation, in contrast to graphene or TMDs that are sufficiently inert to enable ultrasonic processing in aqueous surfactant solutions. In particular, surfactant micelles in aqueous solution do not fully exclude solvated $O_2$ or $H_2O$, allowing potential reactions between BP and these oxidizing agents.[18] Furthermore, common dispersants for graphene in organic solvents can also degrade BP. For example, BP nanosheets are amorphized by ethyl cellulose (see Supporting Information, Figure S1), which is a conventional dispersant for highly concentrated graphene dispersions.[39,40]

Herein, we present a scalable method for preparing pristine 2D BP nanosheets *via* direct liquid exfoliation in organic solvents. By employing a sealed tip ultrasonication system, BP is exfoliated into anhydrous, oxygen-free solvents, avoiding the known chemical degradation pathways for 2D BP. The structure, chemistry, and stability of these solvent-exfoliated BP nanosheets are quantified through a comprehensive suite of measurements including atomic force microscopy (AFM), high-resolution transmission electron microscopy (HRTEM), Raman spectroscopy, and X-ray photoelectron spectroscopy (XPS). Finally, field-effect transistors (FETs) are fabricated from individual solvent-exfoliated BP nanosheets to investigate charge transport. By all of these metrics, our solvent-exfoliated 2D BP nanosheets show behavior that is comparable to mechanically exfoliated 2D BP, suggesting their use in a wide range of semiconductor applications.

**Results and Discussion**

Bulk BP crystals are exfoliated in organic solvents using ultrasonication procedures that are detailed in the Methods section. Briefly, all BP solution preparation takes place in a dark Ar glove box using a modified sealed-tip ultrasonicator setup, as shown in Figs. 1a-b. The sealed container lid is connected to a 0.125" ultrasonicator tip, which is driven at a higher power than typical bath sonication powers in order to minimize the ultrasonication duration (Figs. 1a-b). Specifically, ultrasonication in an ice bath at ~30 W power achieves a BP concentration of ~1 mg/mL in 1 hr in contrast to the 15-24 hrs needed to exfoliate BP *via* bath sonication.[34,35] This reduction in BP sonication time, combined with our anhydrous and anoxic protocol, ultimately preserves the chemical integrity of the BP as will be delineated below. The resulting stable BP dispersions possess a brown to yellow color, as seen in Figure 1c.

Previous graphene studies have found that the choice of organic solvent is critical for efficient solvent exfoliation, with NMP and dimethylformamide (DMF) working particularly well due to their relatively high boiling points and surface tension (~40 mJ/m$^2$).[32, 41] To determine the optimal solvent for BP exfoliation, BP crystals were ultrasonicated in acetone, chloroform, hexane, ethanol, isopropanol (IPA), DMF, and NMP. Under identical preparation conditions, these anhydrous solvents were opened only in an Ar glove box to minimize $O_2$ and $H_2O$ contamination. The as-prepared dispersions were then centrifuged at 500-15,000 rpm for 10 min in an effort to isolate well-exfoliated 2D BP nanosheets, ultimately resulting in the solution color evolving from brown to yellow (Figure 1c). The optical absorbance per cell length ($A/l$) of the BP dispersions before and after centrifugation is measured at 660 nm. The concentration ($C_{BP}$) is then determined based on an extinction coefficient ($\alpha$) of 267 ± 23 Lg$^{-1}$m$^{-1}$ (Figure S2). Figures 1d-e show the BP concentration values for the seven aforementioned solvents, based on their boiling points and surface tension ($A/l$ at 660 nm are indicated in Table S1). The BP concentration monotonically increases with increasing boiling point and surface tension, which qualitatively matches the trend observed for graphene.[41] Based on these results, NMP was identified as the optimal solvent to achieve stable BP dispersions.

Following exfoliation in NMP, the BP nanosheets were characterized with a comprehensive suite of microscopy and spectroscopy methods. For AFM and SEM analysis, samples were prepared in an Ar glove box by drop casting the BP dispersion onto 300 nm SiO$_2$/Si substrates. In particular, a drop of the BP dispersion was placed on the substrate for ~5 min, after which the sample was blown off with $N_2$ gas and annealed on a ~70 °C hotplate for ~2 min. In Figure 2a, the deposited BP is imaged with AFM, employing an environmental cell with dry, ultrahigh purity $N_2$ gas. The AFM height image shows that the thickness values range from 16 nm to 128 nm. No bubbles, droplets, or other signs of BP degradation are present (Figure 2a).[18] Further AFM studies based on centrifugation speed and ambient degradation will be discussed later. The SEM image of Figure 2b indicates that the lateral dimensions of the BP flakes agree well with the AFM images.

For additional atomic-scale characterization, solvent-exfoliated BP flakes were deposited onto holey carbon TEM grids for HRTEM analysis. Figure 2c shows a low-resolution TEM image of a representative BP nanosheet. A schematic of the BP crystal structure is provided in Figure 2d, revealing the A and B high symmetry directions. The HRTEM images of Figures 2e and 2f provide perspectives of these A and B directions, respectively, which are consistent with the inset BP crystal structures. Furthermore, the BP orthorhombic crystalline character is confirmed by selected area electron diffraction (SAED) pattern, as seen in Figure 2g. Overall, this HRTEM analysis provides strong evidence that BP nanosheets exfoliated by the tip sonication maintain their crystalline nature. This conclusion is corroborated by Raman spectroscopy, as given in Figure 2h. Specifically, four modes are observed at ~362 $cm^{-1}$, ~439 $cm^{-1}$, ~467 $cm^{-1}$, and ~521 $cm^{-1}$, which correspond to the $A_g^1$, $B_{2g}$, and $A_g^2$ phonon modes for BP and the TO phonon for the silicon substrate (Figure 2h), respectively.[4] The sharp Lorentzian lineshapes for the $A_g^1$, $B_{2g}$, and $A_g^2$ modes (1.9 $cm^{-1}$, 3.7 $cm^{-1}$, and 2.6 $cm^{-1}$, respectively) are consistent with crystalline BP nanosheets.

The chemical quality of the solvent-exfoliated BP flakes is assessed using XPS in Figure 3. The BP samples for XPS analysis are prepared using the same procedure as described for Figure

2. For reference, a BP bulk crystal (from HQ Graphene) is also measured. The solvent-exfoliated BP nanosheets are monitored as a function of ambient exposure time, in order to facilitate comparison with mechanically exfoliated BP, whose ambient oxidization kinetics are known.[18] Figure 3a shows that the as-prepared BP flakes have the $2p^{3/2}$ and $2p^{1/2}$ doublet at 129.7 eV and 130.5 eV, respectively, characteristic of crystalline BP.[18, 42-44] Small, oxidized phosphorus (*i.e.*, $PO_x$) subbands are also apparent at ~134 eV, as observed in previous measurements.[18, 42-44] These $PO_x$ subbands likely stem from oxygen defects[45] or surface suboxides in the BP, which are introduced during solvent exfoliation. Nevertheless, these defects do not significantly compromise the electronic characteristics of the BP nanosheets, as will be detailed later. Figures 3b-e reveal that, despite 1, 2, 3, and 7 days of ambient exposure, the BP nanosheets have a similar oxide content, demonstrating slowed ambient degradation kinetics for solvent-exfoliated BP nanosheets relative to mechanically exfoliated BP.[18]

In Figures 3f and 3g, *in situ* sputtering is employed to remove the obscuring oxygen defects in both the solvent-exfoliated BP nanosheets and bulk BP crystals. In this manner, the solvent-exfoliated BP nanosheets can be compared against a chemically, electronically, and structurally pure standard. The XPS spectra are taken after sputtering the solvent-exfoliated BP sample for 10 s (*ca.* 2.4 nm BP material removed) and the bulk BP crystal for 60 s (*ca.* 14.4 nm removed). Figures 3f and 3g show P 2p core level XPS data for these two cases. Both spectra have weak $PO_x$ subbands after sputtering, resulting from trace amounts of oxygen defects in the samples. When compared against the BP bulk crystal, the solvent-exfoliated BP sample has a lower XPS signal intensity, due to a lower amount of BP material present. Other than the signal intensity, however, the two spectra are indistinguishable, further reinforcing the high chemical quality of the solvent-exfoliated BP nanosheets.

In order to tailor the size distribution of solvent-exfoliated BP nanosheets, the dispersed BP solutions are centrifuged at different speeds. In Figure 4a, the solution color proceeds from dark brown to light yellow as higher centrifugation speeds are applied. Figure 4b reveals the correlation between centrifugation speed and overall BP concentration, with the light yellow solution ("5") possessing the most dilute concentration (~0.01 mg/mL) of BP nanosheets. In addition, the flake thickness and lateral size decrease with increasing centrifugation, as determined by AFM and shown in Figure 4c. The BP dispersions centrifuged at 500 rpm possess a significant number of thick BP nanosheets (>50 nm thick BP outliers are indicated in Figure S3). In contrast, BP dispersions following 10,000 rpm and 15,000 rpm centrifugation minimize the BP nanosheet lateral size. Finally, BP dispersions centrifuged at 5,000 rpm result in thin BP flakes with relatively large lateral area, which will be used for the charge transport measurements described below.

To determine suitable processing conditions for solvent-exfoliated BP devices, an ambient degradation study was performed using AFM. Recently, Wood *et al*. reported that topographic protrusions (hereafter called bubbles) were found within mechanically exfoliated BP nanosheets shortly after exfoliation.[18] Structural modifications such as bubble formation in BP flakes are an indication of irreversible BP oxidation into phosphate derivatives, ultimately destroying the high-performance electronic properties of BP. Five different samples were prepared on identical 300 nm $SiO_2$/Si substrates to assess potential solvent-exfoliated BP structural modification in ambient conditions. Figure 5 shows AFM amplitude images for BP samples prepared by mechanical exfoliation, liquid exfoliation in NMP, and mechanical exfoliation with 1 hr dipping in NMP. In addition, Figure S6 gives further AFM data for atomic layer deposition (ALD) alumina-encapsulated samples produced by solvent exfoliation in NMP and mechanical exfoliation. After sample preparation, all samples were stored in dark, ambient conditions with an average temperature of 26.9 ± 0.2 °C and a relative humidity of 21.2 ± 0.4%.

In Figure 5, no bubbles or other evidence of degradation are observed for all BP samples shortly after sample preparation (red arrows). After 1 day in ambient conditions, bubbles (blue arrows) are apparent only on the mechanically exfoliated BP surface, corroborating recently reported results.[18] After 2 days, bubbles are also found on the solvent-exfoliated BP sample and the mechanically exfoliated BP sample with 1 hr NMP exposure. These degradation bubbles occur after 2 days, regardless of flake thickness or lateral size (see Figures S4 and S5). After 7 days, we note that the bubbles coarsen, forming larger and taller bubbles on all three samples. Here, the red and blue arrows indicate the same position before and after the appearance of bubbles. From these results, it appears that residual NMP retards BP degradation for about 24 hrs of total ambient exposure, potentially originating from NMP encapsulation or intercalation. Control AFM measurements on solvent-exfoliated and mechanically exfoliated BP samples passivated with ALD alumina overlayers do not show any degradation bubbles after 7 days, as expected (Figure S6). Overall, these results underscore that solvent-exfoliated BP nanosheets possess comparable, if not slightly improved, ambient stability, thus enabling reliable device fabrication and charge transport measurements.

To explore the electrical properties of solvent-exfoliated BP nanosheets, FET devices were fabricated on individual BP flakes by electron beam lithography (EBL) using metal electrodes consisting of 20 nm Ni and 40 nm Au, as shown in Figure 6a. To enable high device yield, dispersions were prepared with BP nanosheet lateral dimensions suitable for EBL processing and thicknesses ($t$) that are expected to demonstrate semiconducting behavior ($t < 10$ nm). Based upon the lateral size and thickness histograms (Figures 4b and 4c), BP dispersions were centrifuged at 5000 rpm to achieve a distribution of BP nanosheets with proper thickness and lateral size. To further increase the number density of the BP nanosheets, the BP nanosheets were collected on anodic aluminum oxide (AAO) membranes with 20 nm pore size *via* vacuum filtration. The BP flakes on these membranes were then transferred onto 300 nm $SiO_2$/Si substrates using polydimethylsiloxane (PDMS) stamping. From optical microscopy imaging in Figure 5b, the highly packed BP nanosheets were found to have a color contrast of dark green to light yellow (Figure 5b, left). This level of BP nanosheet surface coverage facilitated the fabrication of individual BP nanosheet FETs. However, we note that our protocol may be modified to achieve degradation-free, continuous BP thin films by controlling substrate wettability with surface pre-treatments,[18] by intercalating lithium to increase exfoliation yield,[29] or by employing Langmuir-Blodgett thin film deposition techniques.[46] Although the BP nanosheets were exposed to ambient conditions for ~60 to 90 min during this processing, the results of Figure 5 suggest that degradation on this timescale is minimal. Figure 6b, right, shows the as-fabricated, solvent-exfoliated BP FET.

FET measurements reveal ambipolar behavior for solvent-exfoliated BP with a median hole mobility of 25.9 $cm^2V^{-1}s^{-1}$ and $I_{on}/I_{off}$ ratio of 1.6 x $10^3$. To demonstrate the reproducibility of these transport results, mobility and $I_{on}/I_{off}$ ratio histograms from as-fabricated, solvent-exfoliated BP FETs are presented in Figures 6c and 6d. The solution-processed BP FETs uniformly show current saturation and ambipolar transfer characteristics. In Figures 6e and 6f, the transfer and output curves are shown for our champion, solution-processed BP FET, having an $I_{on}/I_{off}$ ratio of ~$10^4$. The transfer curve of a BP FET showing both gate voltage sweep directions is shown in Figure S7a. The observed level of hysteresis is comparable to previous results on micromechanically exfoliated BP FETs.[18,23] It should also be noted that the field-effect mobility extracted from the forward sweep (75.5 $cm^2$/Vs) is consistently higher than the mobility extracted from the reverse sweep (Figure S7b). Overall, these FET results are comparable to micromechanically exfoliated BP, further indicating that our solvent-exfoliation method does not compromise electrical or structural properties.

**Conclusion**

In summary, a solvent-based exfoliation method is presented that produces electronic-grade BP nanosheets using a sealed-tip ultrasonicator at high power output in an inert environment. The solvent-exfoliated BP nanosheets are then characterized with a comprehensive set of microscopic and spectroscopic analysis techniques, revealing structural and chemical properties that are comparable to mechanically exfoliated BP. Furthermore, AFM measurements as a function of ambient exposure show that residual NMP provides a reduction in the ambient degradation kinetics for solvent-exfoliated BP compared to mechanically exfoliated BP. This stability coupled with the appropriate centrifugal size selection of solvent-exfoliated BP nanosheets allows for the reliable preparation of FET devices. The resulting charge transport measurements show that the solvent-exfoliated BP nanosheets possess ambipolar behavior with device metrics that are competitive with mechanically exfoliated BP. All of our microscopic, spectroscopic, and electronic transport measurements demonstrate that our liquid-phase exfoliated BP nanosheets rival the characteristics of pristine, mechanically exfoliated BP flakes. The ability to produce electronic-grade BP nanosheets *via* scalable solution processing will accelerate ongoing efforts to realize large-area 2D BP electronic and optoelectronic applications.

**Methods**

**Solvent exfoliation.** Black phosphorus (BP) crystals were purchased from a commercial supplier (Smart-Elements) and stored in a dark Ar glove box. All BP crystal storage containers were annealed in the Ar glove box (>100 °C) to remove adventitious $H_2O$. For solvent exfoliation experiments, a custom tip sonicator setup was prepared by perforating the plastic lid of a 50 mL conical tube with a 0.125" sonicator tip. The interface between the tip and the lid was sealed with PDMS several times, preventing $O_2$ and $H_2O$ penetration into the tube (see Figures 1a,b). Additionally, Parafilm and Teflon tapes were used to further suppress potential $O_2$ or $H_2O$ pathways between the lid and container. Therefore, only BP crystals, anhydrous solvent, and Ar gas were present in the conical tube during sonication. The BP crystal and organic solvent were placed in this sealed conical tube with the sonicator tip in an Ar glove box. The container was then connected to the sonicator (Fisher Scientific Model 500 Sonic Dismembrator) in ambient conditions, after which the BP crystal was exfoliated *via* ultrasonication. As-prepared BP dispersions were centrifuged with different speeds to remove unexfoliated BP crystals using an Eppendorf tabletop centrifuge (Figure 1c).

**Extinction coefficient measurements.** Different volumes of BP dispersion in NMP after centrifugation were vacuum filtered on anodized aluminum oxide (AAO) membranes possessing a ~20 nm pore size. The concentration of dispersed BP was calculated by measuring the weight difference of the AAO membrane before and after vacuum filtration. Furthermore, the optical absorbance per cell length (*A/l*) was determined from optical absorbance spectra at 660 nm (Cary 5000 Spectrophotometer, Agilent Technologies). Using Beer's law ($A/l = \alpha \cdot C_{BP}$), the BP extinction coefficient was extracted from the slope of a plot of *A/l versus* concentration (Figure S2).

**Atomic force microscopy (AFM).** All height and amplitude measurements were performed in tapping mode using an Asylum Cypher AFM with Si cantilevers (~290 kHz resonant frequency). Images were taken in the repulsive phase regime using a minimum of 512 samples per line. The scanning rate was ~0.4 Hz. AFM imaging employing $N_2$ flow was performed in an environmental cell attached to the Cypher ES scanner. After the BP was deposited onto the substrate, the environmental cell was assembled in the Ar glove box and attached to an ultra-high purity grade $N_2$ flowing tube. BP flakes of interest for AFM scanning were identified *in situ* with the built-in optical microscope in the Cypher system. During AFM scanning, $N_2$ was continuously flowed through the cell with the optical microscope light illuminated.

For the AFM ambient degradation study, all samples were prepared in the Ar glove box and transferred onto the normal sample holder exposed to ambient conditions, with fewer than 30 min of exposure to ambient air before selecting BP flakes of interest with the built-in optical microscope and initiating the AFM image scan.

**Scanning electron microscopy (SEM).** SEM images using a secondary electron detector were acquired on a Hitachi SU 8030 FE-SEM. The acceleration voltage was 2 keV, and the beam current was ~10 µA. Images were acquired in fewer than 40 s to mitigate sample charging and electron beam induced deposition of carbon.

**Transmission electron microscopy (TEM).** A droplet of BP dispersion was deposited on a holy carbon TEM grid (Ted-Pella) and fully dried in an Ar glove box. The TEM grid was then loaded in the TEM sample holder with fewer than 5 min of exposure to ambient air. The TEM measurement was performed with a JEOL JEM-2100 TEM at an accelerating voltage of 200 keV.

**Raman spectroscopy.** Raman spectra were obtained using an Acton TriVista Confocal Raman System with an excitation wavelength of 514 nm. Data were collected for 30 s at ~0.1 mW using a 100X objective.

**X-ray photoelectron spectroscopy (XPS).** An ultrahigh vacuum (UHV) Thermo Scientific ESCALAB 250 Xi XPS system was used at a base pressure of $\sim 5\times 10^{-10}$ Torr to gather XPS data. The XPS data had a binding energy resolution of ~0.1 eV using a monochromated Al Kα X-ray source at ~1486.7 eV (~400 μm spot size). Solvent-exfoliated BP thin films on $SiO_2$/Si were charge compensated using a flood gun. No flood gun compensation was used for bulk BP crystals (HQ Graphene, The Netherlands). Bulk BP crystals were in electrical contact with the stage by UHV-compatible, double-sided Cu tape. All core level spectra were the average of 5 scans taken at a 100 ms dwell time using a pass energy of 15 eV. When using charge compensation, all core levels were charge corrected to adventitious carbon at ~284.8 eV. A 3000 keV ion gun (*ca.* 0.24 nm/s etch rate) was used to perform depth profiling. Using the software suite Avantage (Thermo Scientific), all subpeaks were determined in a manner detailed elsewhere.[18] The p core levels for phosphorus and silicon were fitted with doublets.

**Electron beam lithography (EBL).** Features were defined with EBL in PMMA. To make electrical contact to the flake, 20 nm of Cr and 40 nm of Au were used as the contact metals. To minimize BP degradation, fewer than 24 hours elapsed between exfoliation and charge transport measurements.

**Atomic layer deposition (ALD).** ALD of alumina was performed with $H_2O$ and trimethylaluminum (TMA) precursors in a Cambridge NanoTech reactor. For the AFM stability studies detailed in the Supporting Information, ~3 nm of alumina was deposited at room temperature. During the ALD process, pulses of TMA precursor were introduced before those of $H_2O$ in an effort to scavenge any adventitious oxygen or $H_2O$. To form a seed layer and protect BP from potential damage due to high temperature exposure to oxygenated $H_2O$, the first ~3 nm were deposited at room temperature with prolonged purge time for both $H_2O$ and TMA. Subsequently, the remainder of the film was deposited at 150 ºC with a normal purge time. No other seeding layers were used in the processing.

**Charge transport measurements.** Electrical measurements of BP FETs were performed in a Lakeshore CRX 4K probe station at less than $5\times10^{-4}$ Torr pressure at room temperature. Two Keithley Source Meter 2400 units were used to measure the current-voltage characteristics. Equation (1) was used to calculate mobility in Figure 6:

$$\mu_{eff} = \frac{L g_d}{W C_{ox} V_{DS}} \qquad (1)$$

where $\mu_{eff}$ is the field effect mobility, $L$ is the channel length (obtained from optical micrographs), $g_d$ is the transconductance, $W$ is the channel width (obtained from optical micrographs), $C_{ox}$ is the oxide capacitance (11 nF·cm$^{-2}$ for 300 nm thick thermal $SiO_2$), and $V_{DS}$ is the applied source-drain bias.

**Supporting Information**

The supporting information gives additional data and analysis including HRTEM images, extinction coefficient measurements, solvent-exfoliated BP thickness histograms, AFM height images of solvent-exfoliated BP flakes, AFM amplitude images of BP flakes encapsulated by ALD, and BP FET hysteresis results. This document is available free of charge *via* the Internet at http://pubs.acs.org.

**Corresponding Author**

*Correspondence should be addressed to m-hersam@northwestern.edu

**Author Contributions**

J.K., J.D.W., S.A.W., and M.C.H. planned the experiments. J.K. performed solution processing, AFM, Raman spectroscopy, and thin film transfer. J.D.W. prepared mechanically exfoliated BP from bulk crystals and performed XPS. S.A.W. prepared devices by electron beam lithography and measured charge transport. J.-H.L. and J.K. performed the AFM ambient stability study. X.L. collected HRTEM and SAED data. K.-S.C. and J.K. collected AFM data in the environmental cell. The manuscript was written through contributions of all authors. M.C.H. supervised the project.

**Acknowledgments**

Solution processing and Raman spectroscopy were supported by the National Science Foundation (DMR-1006391), AFM and XPS were supported by the Office of Naval Research (N00014-14-1-0669), electron microscopy was supported by the Department of Energy (DE-FG02-09ER16109), and charge transport measurements were supported by the National Science Foundation Materials Research Science and Engineering Center (DMR-1121262). This work made use of the NUANCE Center, which has received support from the NSF MRSEC (DMR-1121262), the State of Illinois, and Northwestern University. S.A.W. was supported under contact FA9550-11-C-0028 from the Department of Defense, Air Force Office of Scientific Research, National Defense Science and Engineering Graduate (NDSEG) Fellowship, 32 CFR 168a. The authors kindly thank Dr. Jinsong Wu and Dr. Shuyou Li for their help with TEM measurements. The authors also acknowledge fruitful discussions with Christopher Ryder.

## References

1. Brown, A.; Rundqvis, S. Refinement of Crystal Structure of Black Phosphorous. *Acta Crystallogr.* **1965**, *19*, 684-685.
2. Jamieson, J. C. Crystal Structures Adopted by Black Phosphorus at High Pressures. *Science* **1963**, *139*, 1291-1292.
3. Fei, R. X.; Yang, L. Strain-Engineering the Anisotropic Electrical Conductance of Few-Layer Black Phosphorus. *Nano Lett.* **2014**, *14*, 2884-2889.
4. Xia, F. N.; Wang, H.; Jia, Y. C., Rediscovering Black Phosphorus as an Anisotropic Layered Material for Optoelectronics and Electronics. *Nat. Commun.* **2014**, *5,* 4458.
5. Novoselov, K. S.; Jiang, D.; Schedin, F.; Booth, T. J.; Khotkevich, V. V.; Morozov, S. V.; Geim, A. K. Two-Dimensional Atomic Crystals. *Proc. Natl. Acad. Sci. U. S. A.* **2005**, *102*, 10451-10453.
6. Green, A. A.; Hersam, M. C. Solution Phase Production of Graphene with Controlled Thickness *via* Density Differentiation. *Nano Lett.* **2009**, *9*, 4031-4036.
7. Jariwala, D.; Sangwan, V. K.; Lauhon, L. J.; Marks, T. J.; Hersam, M. C. Emerging Device Applications for Semiconducting Two-Dimensional Transition Metal Dichalcogenides. *ACS Nano* **2014**, *8*, 1102-1120.
8. Kang, J.; Seo, J. W.; Alducin, D.; Ponce, A.; Yacaman, M. J.; Hersam, M. C. Thickness Sorting of Two-Dimensional Transition Metal Dichalcogenides *via* Copolymer-Assisted Density Gradient Ultracentrifugation. *Nat. Commun.* **2014**, *5*, 5478.
9. Kim, I. S.; Sangwan, V. K.; Jariwala, D.; Wood, J. D.; Park, S.; Chen, K. S.; Shi, F. Y.; Ruiz-Zepeda, F.; Ponce, A.; Jose-Yacaman, M. *et al.* Influence of Stoichiometry on the Optical and Electrical Properties of Chemical Vapor Deposition Derived $MoS_2$. *ACS Nano* **2014**, *8*, 10551-10558.
10. Takao, Y.; Asahina, H.; Morita, A. Electronic-Structure of Black Phosphorus in Tight-Binding Approach. *J. Phys. Soc. Jpn.* **1981**, *50*, 3362-3369.
11. Akahama, Y.; Endo, S.; Narita, S. Electrical-Properties of Black Phosphorus Single-Crystals. *J. Phys. Soc. Jpn.* **1983**, *52*, 2148-2155.
12. Liu, H.; Neal, A. T.; Zhu, Z.; Luo, Z.; Xu, X. F.; Tomanek, D.; Ye, P. D. D. Phosphorene: An Unexplored 2D Semiconductor with a High Hole Mobility. *ACS Nano* **2014**, *8*, 4033-4041.
13. Dai, J.; Zeng, X. C. Bilayer Phosphorene: Effect of Stacking Order on Bandgap and Its Potential Applications in Thin-Film Solar Cells. *J. Phys. Chem. Lett.* **2014**, *5*, 1289-1293.
14. Tran, V.; Soklaski, R.; Liang, Y. F.; Yang, L. Layer-Controlled Band Gap and Anisotropic Excitons in Few-Layer Black Phosphorus. *Phys. Rev. B* **2014**, *89*, 235319.
15. Li, L. K.; Yu, Y. J.; Ye, G. J.; Ge, Q. Q.; Ou, X. D.; Wu, H.; Feng, D. L.; Chen, X. H.; Zhang, Y. B. Black Phosphorus Field-Effect Transistors. *Nat. Nanotechnol.* **2014**, *9*, 372-377.
16. Appalakondaiah, S.; Vaitheeswaran, G.; Lebegue, S.; Christensen, N. E.; Svane, A. Effect of Van Der Waals Interactions on the Structural and Elastic Properties of Black Phosphorus. *Phys. Rev. B* **2012**, *86*, 035105.
17. Keyes, R. W. The Electrical Properties of Black Phosphorus. *Phys. Rev.* **1953**, *92*, 580-584.
18. Wood, J. D.; Wells, S. A.; Jariwala, D.; Chen, K. S.; Cho, E.; Sangwan, V. K.; Liu, X.; Lauhon, L. J.; Marks, T. J.; Hersam, M. C. Effective Passivation of Exfoliated Black Phosphorus Transistors against Ambient Degradation. *Nano Lett.* **2014**, *14*, 6964-6970.

**FIGURES**

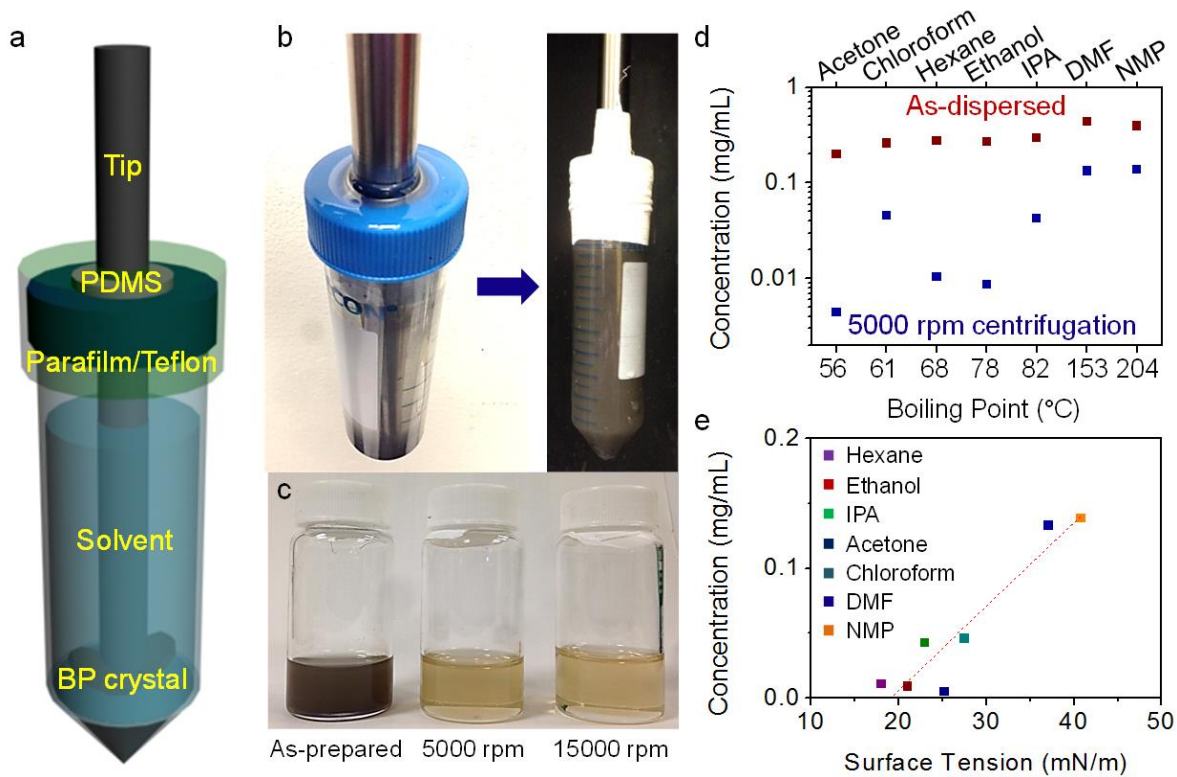

**Figure 1.** Solvent exfoliation of BP in various solvents *via* tip ultrasonication. (a) A schematic and (b) photograph of the custom tip ultrasonication setup that minimized exposure to ambient air during processing. (c) Photograph of a BP dispersion in NMP after ultrasonication, 5000 rpm centrifugation, and 15,000 rpm centrifugation (left to right). (d) BP concentration plot for various solvents with different boiling points before and after 5,000 rpm centrifugation and (e) with different surface tensions after 5,000 rpm centrifugation.

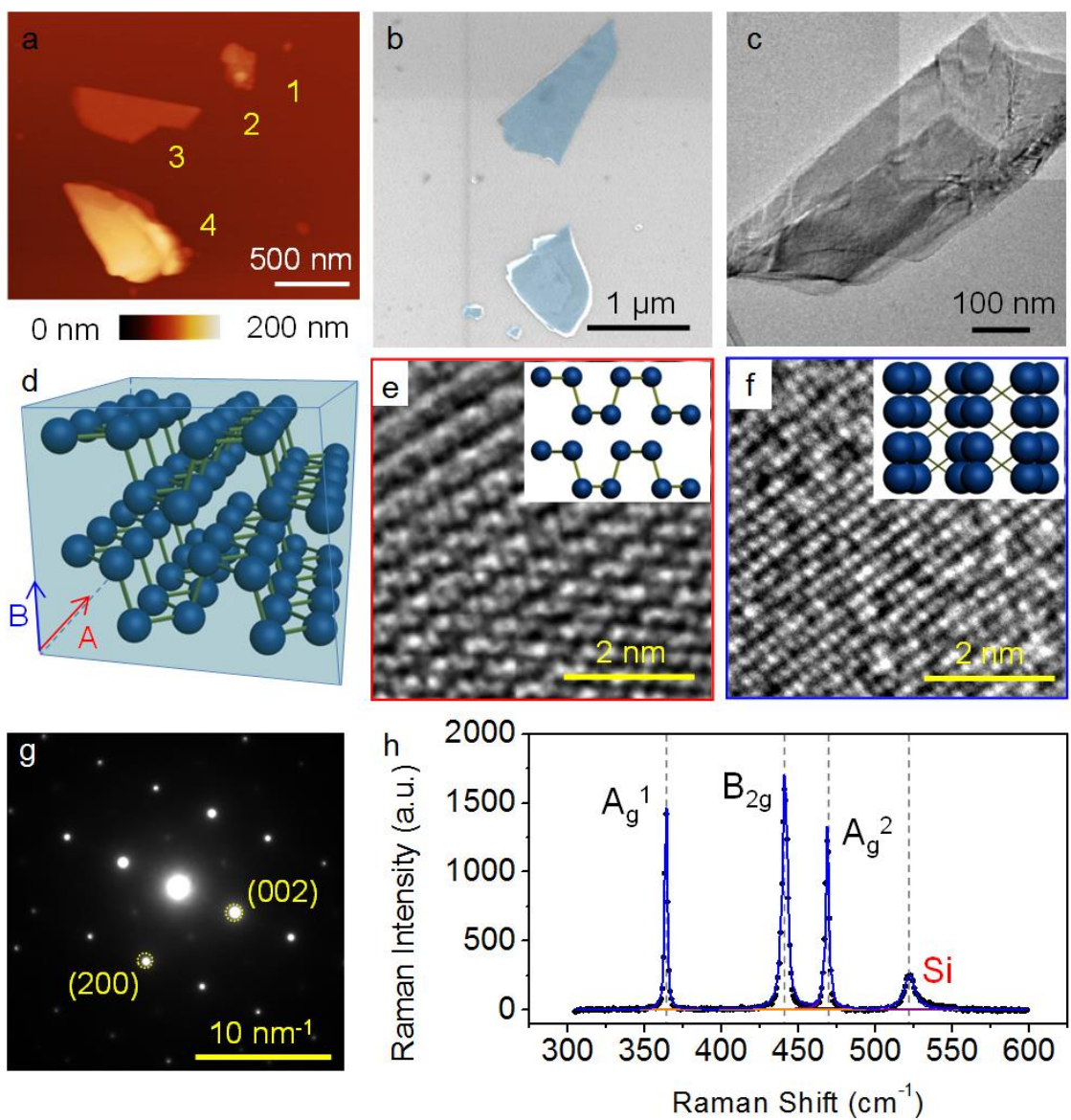

**Figure 2.** Characterization of solvent-exfoliated BP nanosheets. (a) Atomic force microscopy (AFM) height image of solvent-exfoliated BP nanosheets that were deposited onto a 300 nm SiO$_2$/Si substrate with different heights (1: 16 nm, 2: 40 nm, 3: 29 nm, and 4: 128 nm). No bubbles or other evidence of degradation is apparent from the solvent exfoliation process. The height data was taken in a N$_2$ environment. (b) A false-colored scanning electron microscopy (SEM) image of solvent-exfoliated BP nanosheets. (c) Low-resolution transmission electron microscopy (TEM) image of solvent-exfoliated BP nanosheets. (d) Schematic showing the atomic structure of BP. High-resolution TEM images of solvent-exfoliated BP nanosheets along direction (e) A and (f) B. (g) Selected area electron diffraction (SAED) pattern of solvent-exfoliated BP nanosheets. (h) Raman spectrum of solvent-exfoliated BP nanosheets.

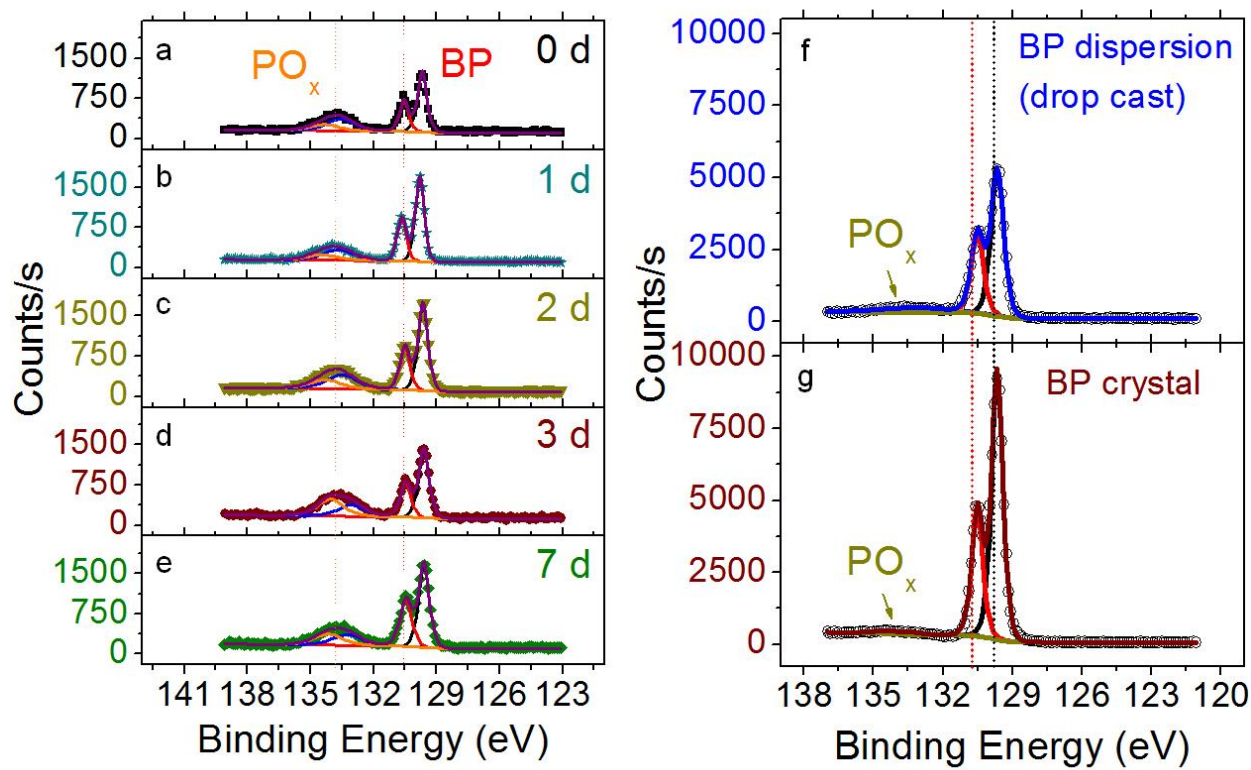

**Figure 3.** Spectroscopic analysis of ambient-aged BP dispersions *versus* bulk BP crystals. P 2p core level X-ray photoelectron spectra (XPS) for drop-casted BP dispersions after (a) 0 days, (b) 1 day, (c) 2 days, (d) 3 days, and (e) 7 days in ambient conditions. The BP flakes are partly oxidized, as confirmed by the $PO_x$ peaks. Oxide content does not increase with exposure time. P 2p core level spectra for (f) drop-casted BP flakes from dispersion and (g) bulk BP crystals. Both spectra are taken after 10 sec of *in situ* sputtering, which removes most of the oxidized BP present.

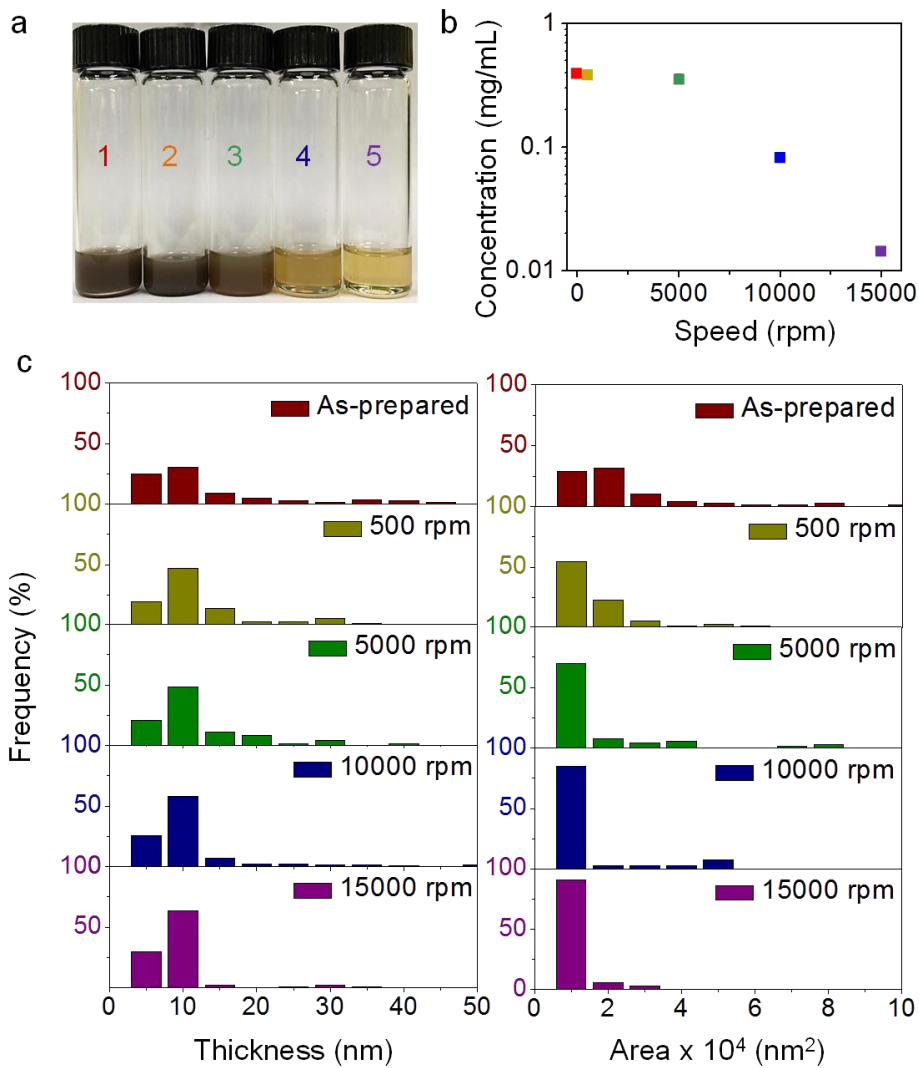

**Figure 4.** Concentration, thickness, and lateral area distribution of solvent-exfoliated BP with different centrifugation speeds. (a) A photograph of BP dispersions in NMP following different centrifugation conditions (1: As-prepared, 2: 500 rpm, 3: 5,000 rpm, 4: 10,000 rpm, and 5: 15,000 rpm). (b) Concentration of the five BP dispersions from part (a). (c) Thickness and lateral area histograms of the five BP dispersions from part (a) as obtained from AFM images.

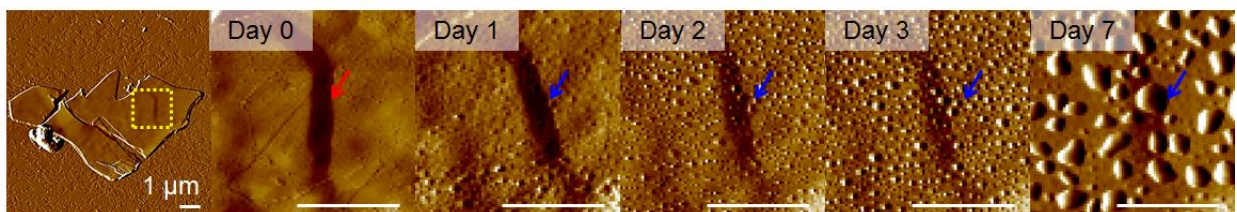
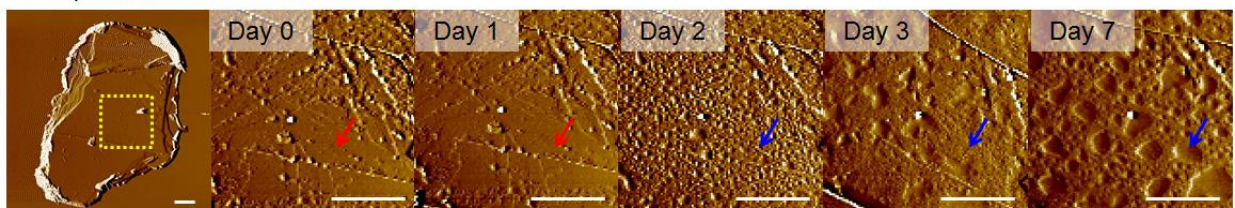
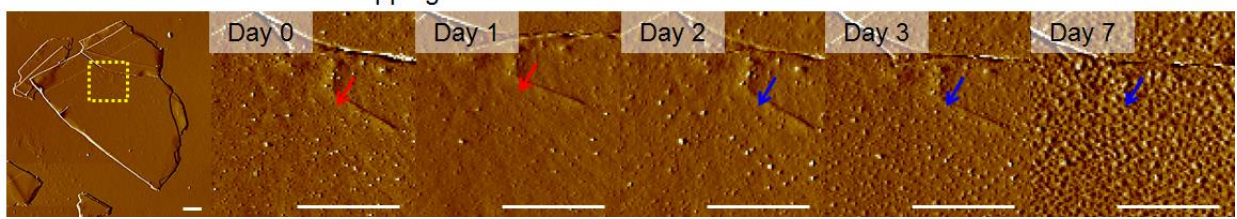

**Figure 5.** AFM amplitude images (amplitude scale: -5 to 5 nm (top left), −1 to 1 nm (magnified images)) of BP flakes prepared by (a) mechanical exfoliation, (b) solvent exfoliation in NMP, and (c) mechanical exfoliation followed by 1 hr submersion in NMP. The leftmost image shows the entire flake, and the images progressing to the right show magnified views immediately after exfoliation up to 7 days in ambient conditions. Structural deformations (*i.e.,* apparent bubbles) are observable on the mechanically exfoliated sample after 1 day and on the rest of samples after 2 days. Red and blue arrows indicate the same position on the BP flake before and after the appearance of bubbles, respectively. All flakes are thicker than 150 nm, and all scale bars are 1 µm.

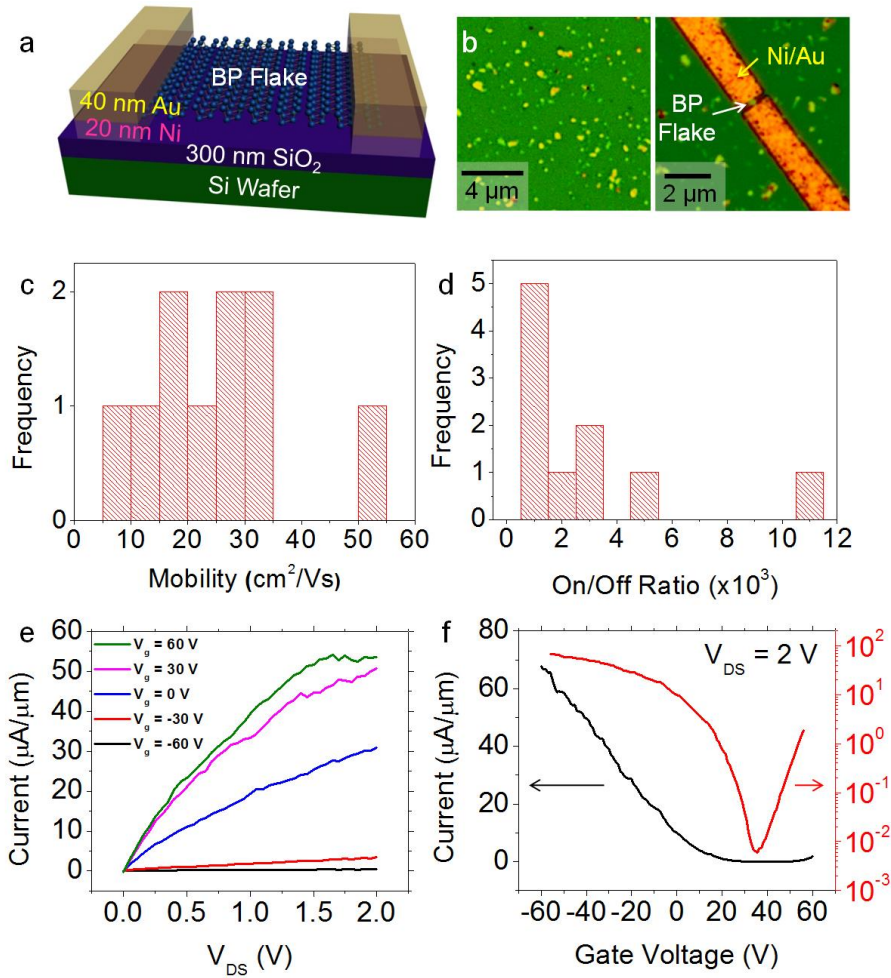

**Figure 6.** Charge transport measurements of solvent-exfoliated BP field-effect transistors (FETs). (a) BP FET device schematic. (b) Optical microscopy images after thin film transfer using PDMS stamping (left) and after device fabrication (right). Histograms of (c) mobility and (d) on/off ratio for a series of solvent-exfoliated BP FETs. (e) Output curves for a solvent-exfoliated BP FET. (f) Transfer curve for a solvent-exfoliated BP FET. The drain current is indicated as a function of gate voltage on a linear scale (black, left) and a logarithmic scale (red, right).

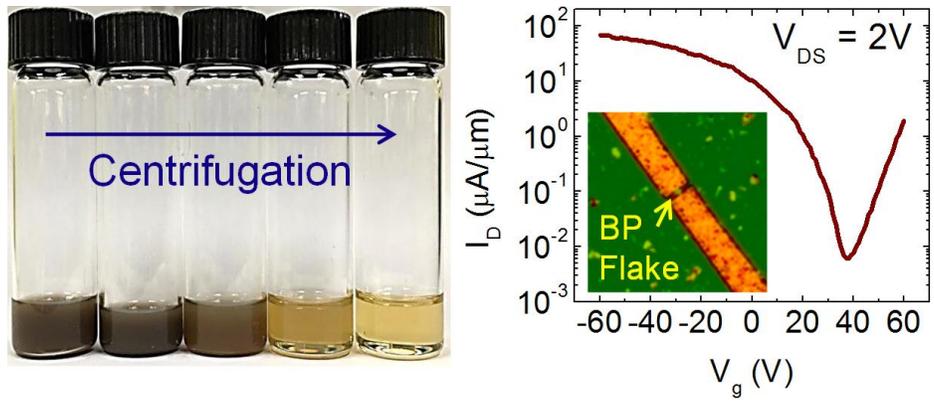

**TOC Figure**

# Supporting Information

# Solvent Exfoliation of Electronic-Grade, Two-Dimensional Black Phosphorus

*Joohoon Kang[1], Joshua D. Wood[1], Spencer A. Wells[1], Jae-Hyeok Lee[1], Xiaolong Liu[2], Kan-Sheng Chen[1], and Mark C. Hersam[1,2,3,4]\**

[1]Department of Materials Science and Engineering, Northwestern University, Evanston, IL 60208
[2]Graduate Program in Applied Physics, Northwestern University, Evanston, IL 60208
[3]Department of Chemistry, Northwestern University, Evanston, IL 60208
[4]Department of Medicine, Northwestern University, Evanston, IL 60208
\* Correspondence should be addressed to m-hersam@northwestern.edu

**Contents:**

- **Table S1.** Absorbance per length at 660 nm

- **Figure S1.** High-resolution TEM images of solvent-exfoliated BP nanosheets

- **Figure S2.** Extinction coefficient of BP in NMP

- **Figure S3.** Thickness histograms of BP dispersions prepared with different centrifugation speeds

- **Figure S4.** AFM height images of BP flakes as a function of ambient exposure

- **Figure S5.** Additional AFM height images of solvent-exfoliated BP flakes as a function of ambient exposure

- **Figure S6.** AFM amplitude images of BP flakes encapsulated by atomic layer deposition (ALD) as a function of ambient exposure

- **Figure S7**. Hysteresis measurements on BP FETs

**Table S1.** Absorbance per length (x $10^{-4}$ [m$^{-1}$]) at 660 nm

|  | Acetone | Chloroform | Hexane | Ethanol | IPA | DMF | NMP |
|---|---|---|---|---|---|---|---|
| **As-prepared** | 7.5 | 9.8 | 10.3 | 10.1 | 11.1 | 16.4 | 14.7 |
| **5000 rpm** | 0.17 | 1.7 | 0.39 | 0.32 | 1.6 | 5.0 | 5.2 |

**Supporting Figures**

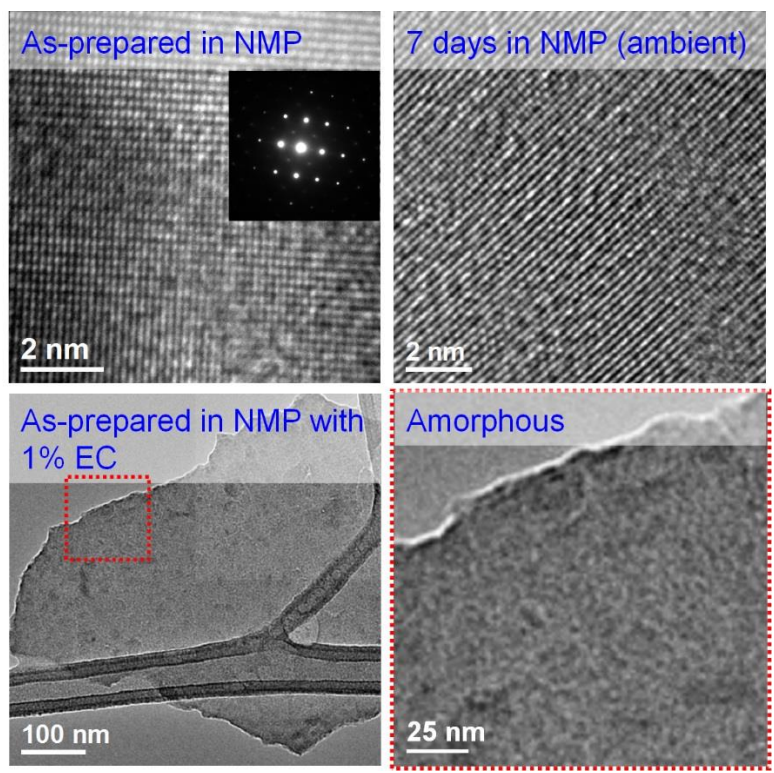

**Figure S1.** High-resolution TEM images of solvent-exfoliated BP nanosheets. Solvent-exfoliated BP nanosheets have an orthorhombic crystalline structure (inset: SAED pattern) for the as-prepared sample (top left) and also for the sample stored in NMP for 7 days in ambient air (top right). Solvent-exfoliated BP nanosheets are fully amorphized after exposure to 1% w/v ethyl cellulose (EC) (bottom left). The bottom right image is a 4x magnified area of the amorphous BP nanosheet.

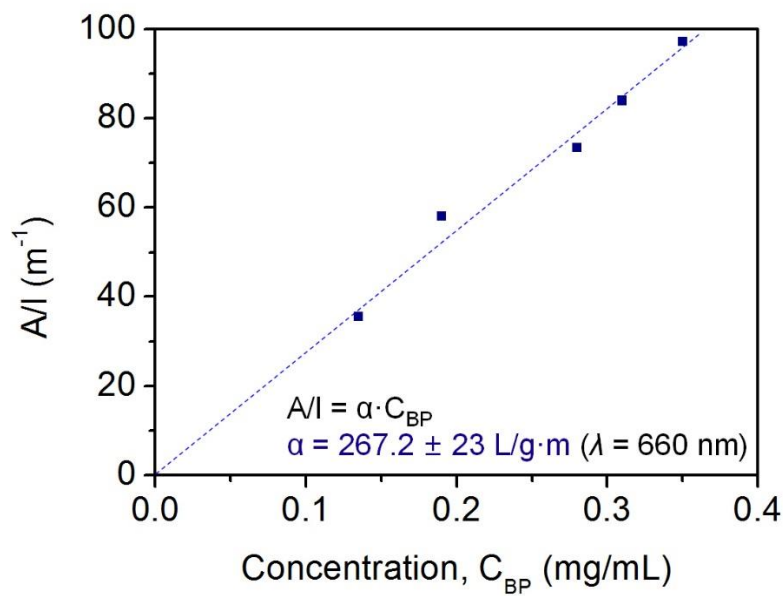

**Figure S2.** The extinction coefficient of BP in NMP is found to be $267 \pm 23$ $Lg^{-1}m^{-1}$ by plotting the absorbance per length (*A/l*) at 660 nm *versus* the concentration of BP.

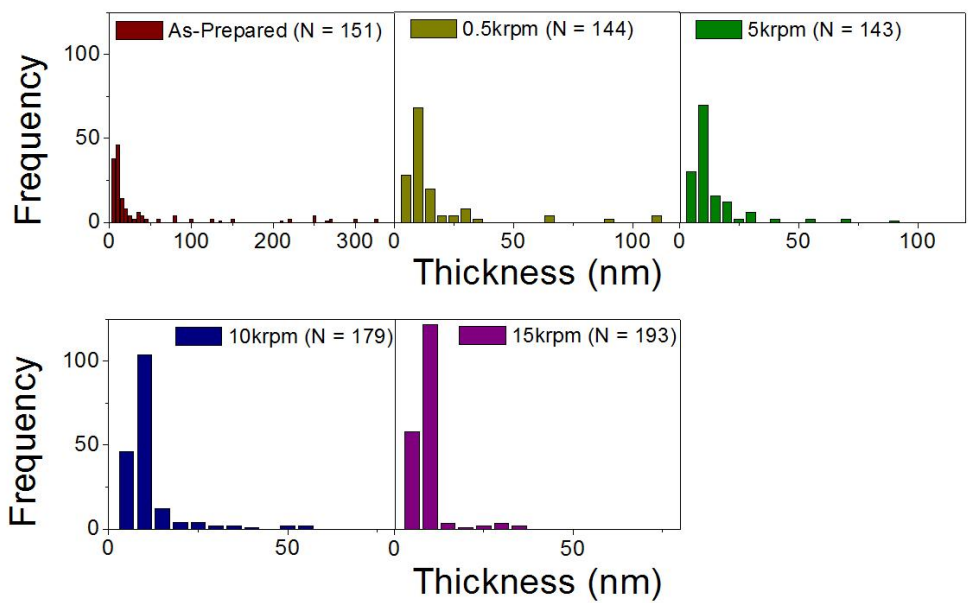

**Figure S3.** Thickness histograms of BP dispersions prepared with different centrifugation speeds. These histograms include outliers with thickness values larger than 100 nm. The thickness values were determined from AFM images.

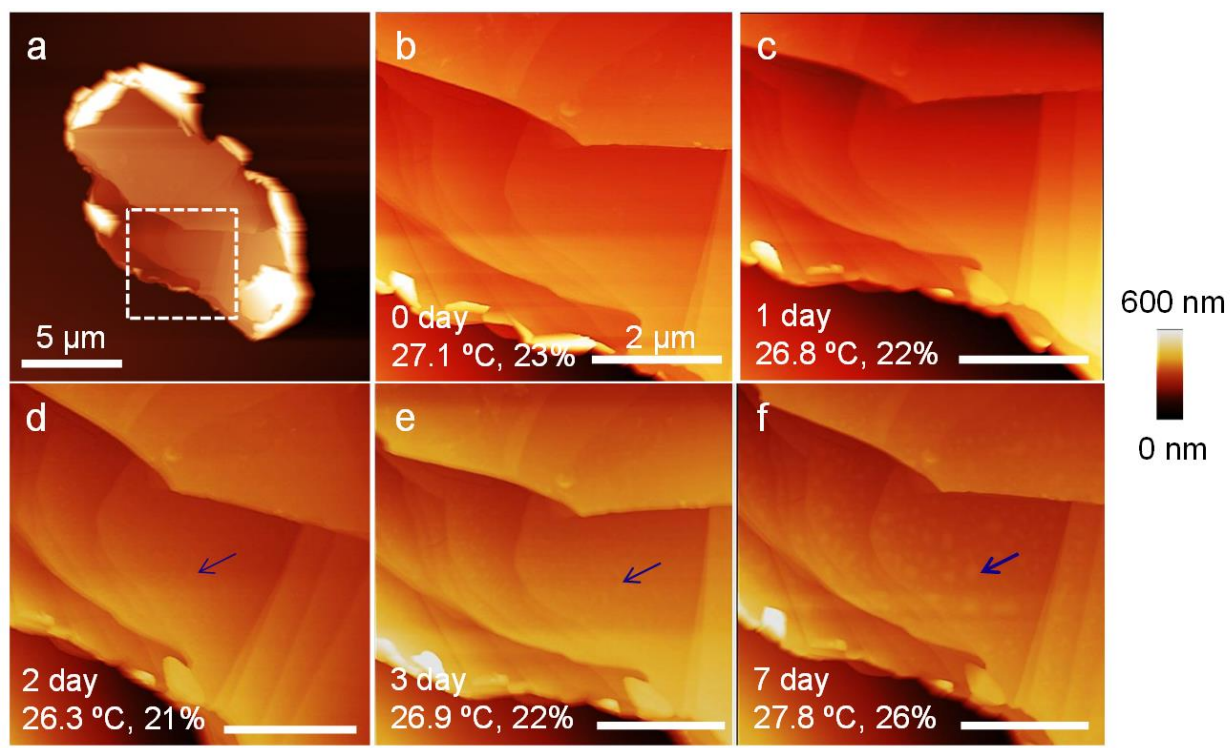

**Figure S4.** AFM height images of a solvent-exfoliated BP flake as a function of ambient exposure (ambient temperature and relative humidity is specified on each image). The BP flake thickness is 266 nm, and all scale bars are 5 μm.

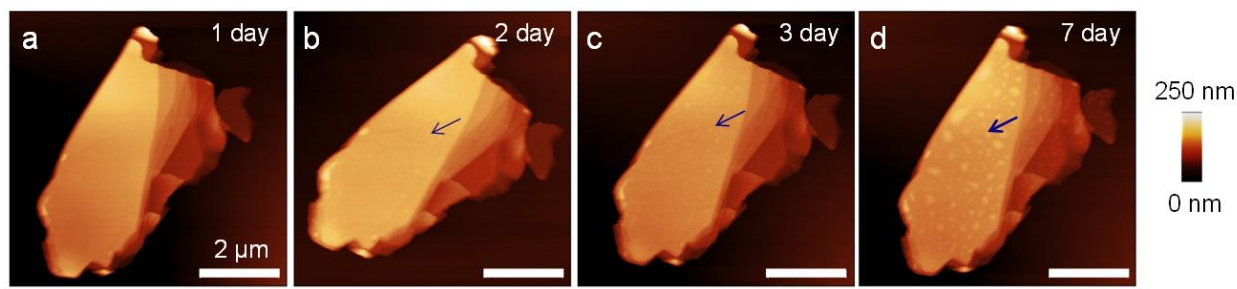

**Figure S5.** Additional AFM height images of a solvent-exfoliated BP flake as a function of ambient exposure. The BP flake thickness is 195 nm, and all scale bars are 2 μm.

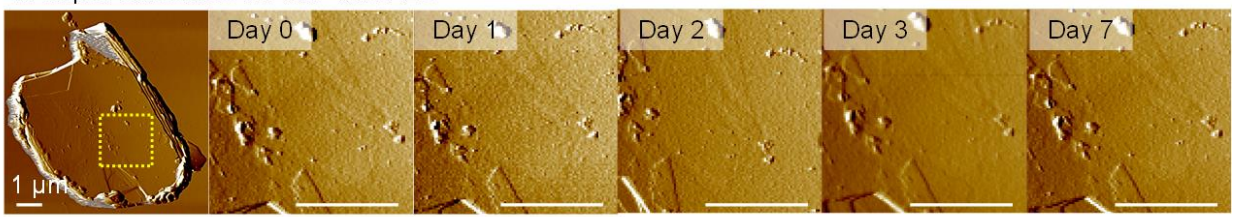

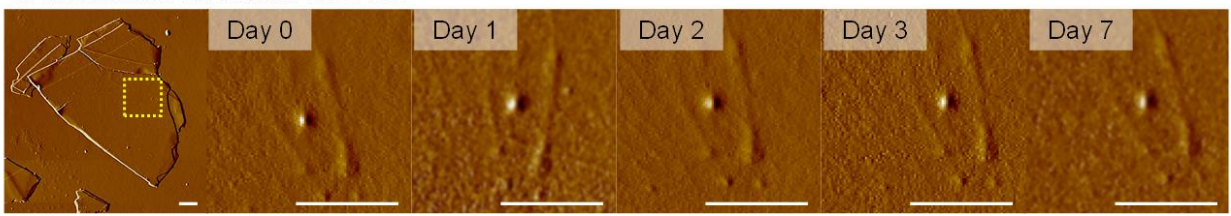

**Figure S6.** AFM amplitude images of BP flakes encapsulated by atomic layer deposition (ALD) alumina as a function of ambient exposure. **(a)** Ambient exposure progression for an ALD-encapsulated, solvent-exfoliated BP flake. **(b)** Ambient exposure progression of an ALD-encapsulated, mechanically exfoliated BP flake. The leftmost image in (a) and (b) corresponds to the entire flake, whereas the other images correspond to higher magnification scans. Day 0 occurs immediately after solvent exfoliation or mechanical exfoliation. No bubbles or other evidence of degradation are evident for all time points, thus illustrating the effectiveness of ALD encapsulation for mitigating ambient degradation for both solvent-exfoliated and mechanically exfoliated BP flakes. Both flakes are thicker than 150 nm, and all scale bars are 1 μm. The amplitude scale is –5 to 5 nm (leftmost images) and –1 to 1 nm (magnified images).

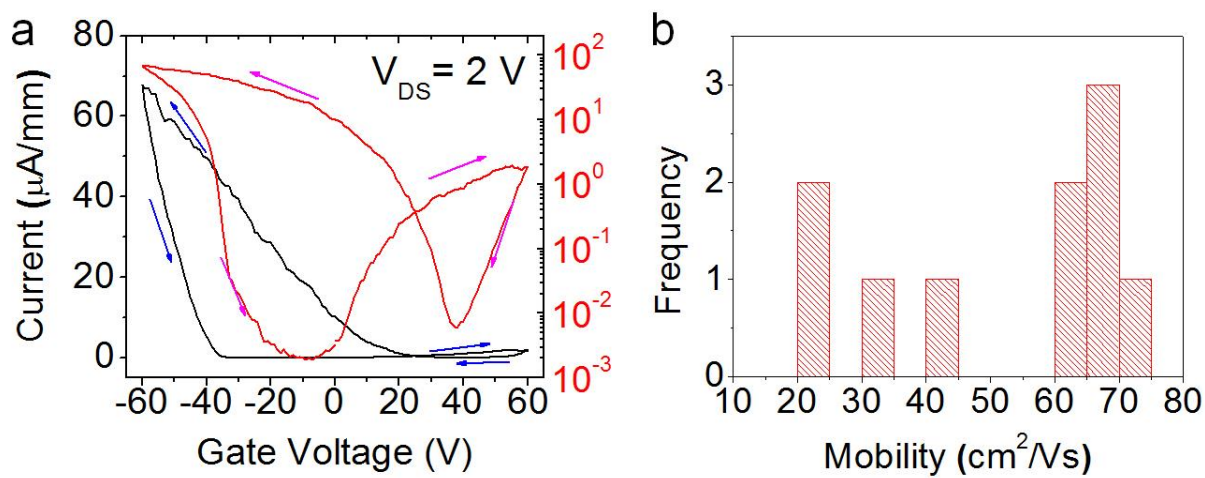

**Figure S7**. Hysteresis measurements on BP FETs. (a) Transfer curve of a solvent-exfoliated BP FET showing both gate voltage sweep directions. The hysteresis is ~40 V, and the field-effect mobility extracted from the forward sweep (75.5 cm$^2$/Vs) is consistently higher than the field-effect mobility extracted from the reverse sweep (25.9 cm$^2$/Vs). The drain current is indicated as a function of gate voltage on a linear scale (black, left) and a logarithmic scale (red, right). (b) Histogram of the field-effect mobility extracted from the forward sweep for several solvent-exfoliated BP FETs.